\documentclass{article}
\usepackage[utf8]{inputenc}
\usepackage[normalem]{ulem}
\usepackage{graphicx} % Include figure files
\usepackage{xcolor}
\usepackage{tablefootnote}
\usepackage{cite}
\usepackage{amsmath,amssymb,amsfonts}
\usepackage{mathtools}
\usepackage{algorithmic}
\usepackage{textcomp}
\usepackage{relsize}

\usepackage[font=small,labelfont=bf]{caption}
\usepackage[square,numbers]{natbib}
\usepackage{hyperref}

\title{Deep Learning Architecture Based Approach For 2D-Simulation of Microwave Plasma Interaction}
\author{Mihir Desai, Pratik Ghosh, Ahlad Kumar and Bhaskar Chaudhury \thanks{Corresponding Author : bhaskar\_chaudhury@daiict.ac.in}\\
\emph{Group in Computational Science and HPC} \\ \emph{DA-IICT, Gandhinagar - 382007 , India.}}
\date{ }
\providecommand{\keywords}[1]{\textbf{\textit{Keywords-}} #1}
\begin{document}
\maketitle
\begin{abstract}
This paper presents a convolutional neural network (CNN)-based deep learning model, inspired from UNet with series of encoder and decoder units with skip connections, for the simulation of microwave-plasma interaction. The microwave propagation characteristics in complex plasma medium pertaining to transmission, absorption and reflection primarily depends on the ratio of electromagnetic (EM) wave frequency and electron plasma frequency, and the plasma density profile. The scattering of a plane EM wave with fixed frequency (1 GHz) and amplitude incident on a plasma medium with different gaussian density profiles (in the range of $1\times 10^{17}-1\times 10^{22}{m^{-3}}$) have been considered. The training data associated with microwave-plasma interaction has been generated using 2D-FDTD (Finite Difference Time Domain) based simulations.  The trained deep learning model is then used to reproduce the scattered electric field values for the 1GHz incident microwave on different plasma profiles with error margin of less than 2\%. We propose a complete deep learning (DL) based pipeline to train, validate and evaluate the model. We compare the results of the network, using various metrics like SSIM index, average percent error and mean square error, with the physical data obtained from well-established FDTD based EM solvers.  To the best of our knowledge, this is the first effort towards exploring a DL based approach for the simulation of complex microwave plasma interaction. The deep learning technique proposed in this work is significantly fast as compared to the existing computational techniques, and can be used as a new, prospective and alternative computational approach for investigating microwave-plasma interaction in a real time scenario.
\end{abstract}

\keywords{
 Deep Learning, CNN, microwave-plasma interaction, FDTD, Computational Electromagnetics, SSIM, multiphysics}
% \textcolor{red}{Mihir all the sections and corresponding subsections are in respective files such as introduction.tex etc}

\section{Introduction}
Microwave-plasma interaction has remained one of the widely researched domain for several decades due to its various applications in electromagnetic (EM) reflectors and absorbers~\cite{vidmar1}, plasma antennas~\cite{alexeff},
plasma stealth technology~\cite{BChaudhury2009RCSopt},
plasma meta-materials~\cite{Sakai2012}, plasma based limiters~\cite{ASemnani2016limit}, switching and protection \cite{semnani,semenov,PratikIMaRC2021,Backstram2011}, plasma diagnostics \cite{shneider}, microwave breakdown applications \cite{bchaudhuryprl,BChaudhury2011, mwbreakdown1}, RF heating of fusion plasmas, and microwave rocket~\cite{mwrocket}. The EM wave propagation into a plasma (complex dispersive medium) primarily depends on the spatial distribution of plasma density and the frequency of the incident EM wave~\cite{BHASKAR2007}. The scattering features of microwaves by an inhomogeneous plasma due to wave reflection and absorption  is usually used to investigate microwave-plasma interaction. In the case of a collisional unmagnetized plasma, the EM wave can propagate in an underdense plasma while being attenuated due to electron-neutral collisions. However, when the plasma density is significantly high (greater than critical density) the EM wave gets reflected and can be completely blocked by an overdense plasma. This interplay of transmission, reflection and absorption of an EM wave propagating in an inhomogeneous plasma leads to complex EM scattering patterns. Several studies have been performed to investigate the effects of different electron density profile on scattering patterns of an incident EM wave of given frequency. The study of microwave-plasma interaction becomes challenging when the plasma is inhomogeneous and the density profile is complicated due to presence of gradients such as a Gaussian density profile.\\
Several theoretical and computational methods exists for the study of EM wave propagation in a plasma \cite{ginzburg,stix,BHASKAR2007}. 
%The magnetoionic theory for the low power EM wave propagation is considered where the governing equations consists of Maxwell’s equations coupled to an auxiliary equation that relates the current and the electric field  \cite{BHASKAR2007}. This auxiliary equation can be derived by solving the equations of motion of the charged particles in the electric field of the EM wave. 
%the wave energy loss due to collisions between the charge particles and the neutral gas atoms and the EM scattering phenomenon that depends on the variation of refractive index  depending on the frequency of the incident EM wave \cite{PratikIMaRC2021}. 
 %The parameter space that takes into account both knowledge of the  plasma (scatterer) and it's EM properties. \\
The FDTD based computational electromagnetics (CEM) technique has remained one of the most preferred methods to model and accurately simulate the microwave-plasma interactions \cite{BHASKAR2007,bchaudhuryieee,konsta,plasmafdtd}. 
%To solve the partial differential equations (PDEs) involved in both the governing Maxwell's equation and the auxiliary equation representing the charged particle motion in the presence of the EM wave, the modelling of the scattering geometry as well as the discretization of the computational domain is one of the important steps in the time domain based simulations \cite{SYan2017}. 
%Different types of meshes such as structured or unstructured have been used for the evaluation of the electric (E) and magnetic (H) fields as well as the charge particle motions. The choice of the mesh depends on the type of computational techniques to evaluate the PDEs. 
%One of the most adopted techniques is the Finite Difference Time Domain (FDTD) \cite{BChaudhury2009RCSopt,BChaudhury2011,PratikIMaRC2021}. Both are time domain techniques where the former generally uses a structured mesh with iterative algorithm to evaluate the EM fields discretized using central difference approximation, and, the later uses a combination of finite volume (FV) and finite element time domain (FETD) to evaluate the EM fields using local unstructured tetrahedral basis on every nodes and weakly maintain a continuity among the nodal volumes (collection of nodes) that are related using the fluxes. The DGTD uses the algebra of matrix representation of the Maxwell's equation and time integration. 
Most of the traditional CEM approaches whether iterative or direct are computationally challenging due to stringent numerical criteria that leads to high memory usage and longer simulation time as the problem size increases \cite{Pratik2020}.
To overcome such challenges inherent to traditional EM solvers, different approaches mostly sought either use of advanced parallelization techniques or selective meshing to reduce long simulation time without loss in accuracy of results \cite{mwcpc,Pratik2020}. High computational cost associated with CEM techniques becomes prohibitive where the real-time analysis of the EM-plasma interaction is of utmost importance. Therefore, it is extremely valuable to explore alternative approaches that can address the problem of high computational cost associated with traditional EM solvers.\\
As described earlier, primarily two sets of data (plasma density profile and scattered EM wave pattern) are associated with microwave-plasma interaction problem, and therefore neural networks can be potentially used to learn the non-linear mappings between these two sets of data and once the network is trained, it can give the outputs in roughly $O(1)$. In the last decade, there has been vast improvements towards the development of large and powerful deep neural networks (DNNs), which has been applied to solve complex problems in the areas of computer vision and image processing. Physics-informed neural networks, a DNN framework, can be also used as a black box to approximate a physical system~\cite{b8} and recent results have shown that DNNs with many layers perform a surprisingly good job in modeling partial differential equation based complicated physics problems in terms of both speed and accuracy~\cite{pnas}. Off late, machine learning (ML)/ deep learning (DL) have also been used to successfully address different complex problems in the areas of plasma physics and computational electromagnetics.
Deep reinforcement learning has been applied for tokamak magnetic controller design to produce new plasma configurations~\cite{nature1}, potential of AI/ML in predicting disruptive instabilities in controlled fusion plasmas has been established in several studies~\cite{nature2,fed}, feasibility of applying ML models for modeling, diagnostics, and control of non-equilibrium plasmas has been discussed in ~\cite{graves} and deep learning has been also used for extracting electron scattering cross sections from plasma swarm Data~\cite{chaudhuryml}.
A Convolutional Neural Networks (CNN) based architecture is learnt to solve full-wave inverse scattering problems~\cite{b12}. The visualizations generated from the problems can be used to train and get the results from the neural network which can potentially help solve and accelerate the traditional equation based solvers \cite{b10}. Deep Learning (DL) as applied to electromagnetics, antenna and EM wave propagation has been well reviewed in \cite{AndreaM2019}. 
%DL has also found its applications in various kinds of plasma modelling problems which deals with Maxwell's equations for electromagnetic waves, Poisson's equations, equations that describe the state of motion, etc. and are often coupled with each other. DL-based Multilayer Perceptron (MLP) and CNN based Particle-in-cell (PIC) methods for plasma simulations has been explored in \cite{b13}.
EM-Net \cite{b3Shutong2020} is an modified end-to-end CNN architecture with residual blocks and skip-connections inspired from the UNet \cite{b2ORonneberger2015}, a robust network with encoder-decoder like structure which generates an image as an output, is widely used in image segmentation problems. \cite{b3Shutong2020} uses EM-Net to predict the electromagnetic field scattered by the complex geometries. \cite{b14} discusses an unsupervised deep learning model which is used for solving time domain electromagnetic simulations, encoding the initial and boundary conditions as well as the Maxwell's equations when training the network. CNNs has been also explored for plasma tomography and disruption prediction from Bolometer data~\cite{plasmatomography}. \cite{poissoneq} compares various CNN based architectures like UNet, MSNet to solve the 2D Poisson equation for electric field computation in Plasma simulations. UNet architecture provides good acceptable results and its capabilities are discussed in the existing literature. This work is aimed at exploring the feasibility of using a CNN based Deep Learning Network, which is heavily inspired from UNet \cite{b2ORonneberger2015}, to accelerate the accurate simulation of microwave-plasma interaction. 

The key contributions of this paper are as follows
\begin{itemize}
    \item To the best of our knowledge, for the first time we propose a deep learning based approach for investigating the interaction of microwaves with an inhomogeneous unmagnetized collisional plasma. 
    \item An end-to-end deep learning model consisting of UNet architecture is used to significantly accelerate the solution of Maxwell's equations coupled with plasma current density term.
        \item A wide range of 2D Gaussian plasma density profiles with different peak electron density values associated with transmission, absorption and reflection of microwaves by the plasma have been considered for generating the training data.
    \item Extensive computational experiments have been carried out to demonstrate the effectiveness of the proposed approach through various ablation studies.

\end{itemize}
   
The remainder of the paper is organized as follows: Section II provides the detailed physics of EM-plasma interaction, the physical model and its numerical implementation. In section III, the proposed deep learning methodology has been discussed. In Section IV, discusses the experimental work, which include the experimental dataset generation, the criteria for parameter selection, the loss function, and subsequently, the simulation results are used to provide effectiveness of the proposed technique by comparing with the existing results and finally, conclusion in section V.

\section{Simulation of microwave plasma interaction}
A substantial number of theoretical, numerical and experimental studies have been carried out to investigate the EM wave propagation characteristics in an unmagnetized collisional plasma. When an EM wave such as microwave is incident on a weakly ionized unmagnetized plasma it is subjected to scattering as well as absorption. The complex relative dielectric permittivity of a collisional plasma
can be expressed as :
\begin{equation} \label{eq1}
    \epsilon (\omega)=\left(1-\frac{\omega_{p}^{2}}{\omega^2+\nu_{m}^{2}}\right)-i\left(\frac{\omega_{p}^{2}}{\omega^2+\nu_{m}^{2}}\right)\left( \frac{\nu_{m}}{\omega}\right)
\end{equation}
The real part in the above equation decides the permittivity and the conductivity is determined by the imaginary part. The conductivity is given by the formula, $\sigma=\epsilon_{0} \nu_{m}\left( \omega_{p}^{2}/(\omega^2+\nu_{m}^{2})\right)$, where $\omega_p=(n_{e} e^{2}/m_e \epsilon_0)^{1/2}$ is the plasma frequency, $\omega$ is the wave angular frequency, $\nu_{m}$ is the
electron-neutral collision frequency, $n_e$ is the local electron density that varies with position, $e$ and $m_e$ represents the electron charge and mass respectively. 
Loss of EM wave energy due to energy transfer to charged particles and subsequently to neutral particles by elastic/ inelastic collisions leads to absorption. Wave scattering is determined by the density variations within the plasma. 
In the case of a microwave whose energy is lower than the ionization potential of the background gas, the wave lacks sufficient energy to further ionize the gas. The plasma acts as a debye dispersive media which responds to the incident EM wave with varying dielectric properties based on the wave frequency and local density.
%The nonhomogeneous plasma with neutrals encounters sufficient collisions when ($\omega \approx \nu$). 
 %\textcolor{red}{you are using different symbols for same quantity at diff locations}
%Based on the electromagnetic theory, the variation in the permittivity ranges, $\epsilon_{r}\leq 1$ to $\epsilon_{r} \leq 0$, that results in transition from real to complex refractive index that decides the wave propagation (either blocked or gets transmitted ) and attenuation (absorption). 
For a fixed wave frequency, the conductivity increases as plasma density increases which results the plasma to behave as a conductor. Coupling of the EM energy to the plasma is decided by the plasma density. At critical density ($n_{critical}$), when the $\omega_p \approx \omega$, the EM wave starts getting reflected. Further, for a collisional plasma, if the plasma density approaches the cutoff density ($n_{cutoff}= n_{critical}(1+(\nu_m/\omega)^{2})$), the plasma shields the incoming microwave resulting in minimum skin depth of EM wave into plasma. 
The plasma density profile controls the different regimes of operation of the plasma from dielectric to a conductor, depending on relation $\omega > \omega_{p}$, $\omega \approx \omega_{p}$ or $\omega < \omega_{p}$, corresponding to transmission, absorption, reflection and minimum penetration into the plasma also referred as skin depth (the distance over which the E-field of the wave decays $1/e$ of it's initial strength). \\
To generate the training data required for ML based approach, we have used the well established EM-plasma fluid model to simulate the physics of EM-plasma interaction, 
 \cite{BChaudhury2010}. The plasma under consideration is a steady-state, nonuniform, cold, weakly ionized, unmagnetized and collisional. %The model follows a Maxwell-Boltzmann Probability Distribution Function to define the plasma species (electrons),  using plasma density, the mean electron velocity and mean energy. 
 The plasma is assumed to be quasi-neutral, and ion contribution to the current density is negligible due to heavier mass of ions.
 The model primarily comprises Maxwell's equations with the electron current density ($J$) term: 

\begin{equation}\label{eq2}
\centering
    \frac{\partial E}{\partial t} \:=\: \frac{1}{\epsilon_0}(\boldsymbol{\nabla}\times H)\:-\: \frac{1}{\epsilon_0}\:(J)
\vspace{-2mm}
\end{equation}
\begin{equation}\label{eq3}
\centering
    \frac{\partial H}{\partial t}\:=\: -\frac{1}{\mu_0}(\boldsymbol{\nabla}\times E)
\vspace{-2mm}
\end{equation}
\begin{equation}\label{eq4}
\centering
    \frac{\partial v_e}{\partial t}\:=\:-\ \frac{e\ E}{m_e}\:-\:\nu_m\ v_e
 \vspace{-1mm}
\end{equation}
where, $\mu_0$ and $\epsilon_0$ represents permeability and electrical permittivity of vacuum respectively, $J$ is the plasma current density $\left(J=-e\:n_e\: v_e\right)$ in (A m$^{-2}$), $e$ is the electron charge ($e=1.602\times10^{-19}$ C), $n_e$ is the electron density in (m$^{-3}$), $v_e$ is the electron velocity in (m/s), $m_{e}$ is mass of electron ($m_{e}=9.1\times10^{-31}$ kg), $\nu_{m}$ is the electron-neutral collision frequency in ($s^{-1})$ (for air plasma considered here, $\nu_m=5.3\times10^{9}\:p$, where $p$ is the ambient pressure in (torr)) \cite{BChaudhury2010}.\\
\indent
The solution to the EM-plasma fluid model can be  numerically achieved by using a FDTD based computational solver that solves the Maxwell's equations and the plasma momentum transfer equation, both are coupled using the current density term $J$. FDTD is an explicit second order accurate time-domain method using centered finite differences on a uniform Cartesian grid, yielding the spatio-temporal
variation of the E and H fields, and has been applied to a wide variety of EM scattering problems \cite{Kunz-1993}. The velocity equation is discretized by the direct integration scheme.
%The solver has been designed for the generation of the actual dataset corresponding to the physics of EM-plasma interaction. The resulting PDEs can be discretized into set of five central difference approximated difference equations that are solved using the FDTD technique. 
%From Eq.\ref{eq:5}(a-e), the grid size along x and y direction ($\Delta_{x}=\Delta_{y}=\Delta$). 

%%%%% Figure1 %%%%
\begin{figure}[h]
\centering
\includegraphics[width=0.4\textwidth, height=0.22\textwidth]{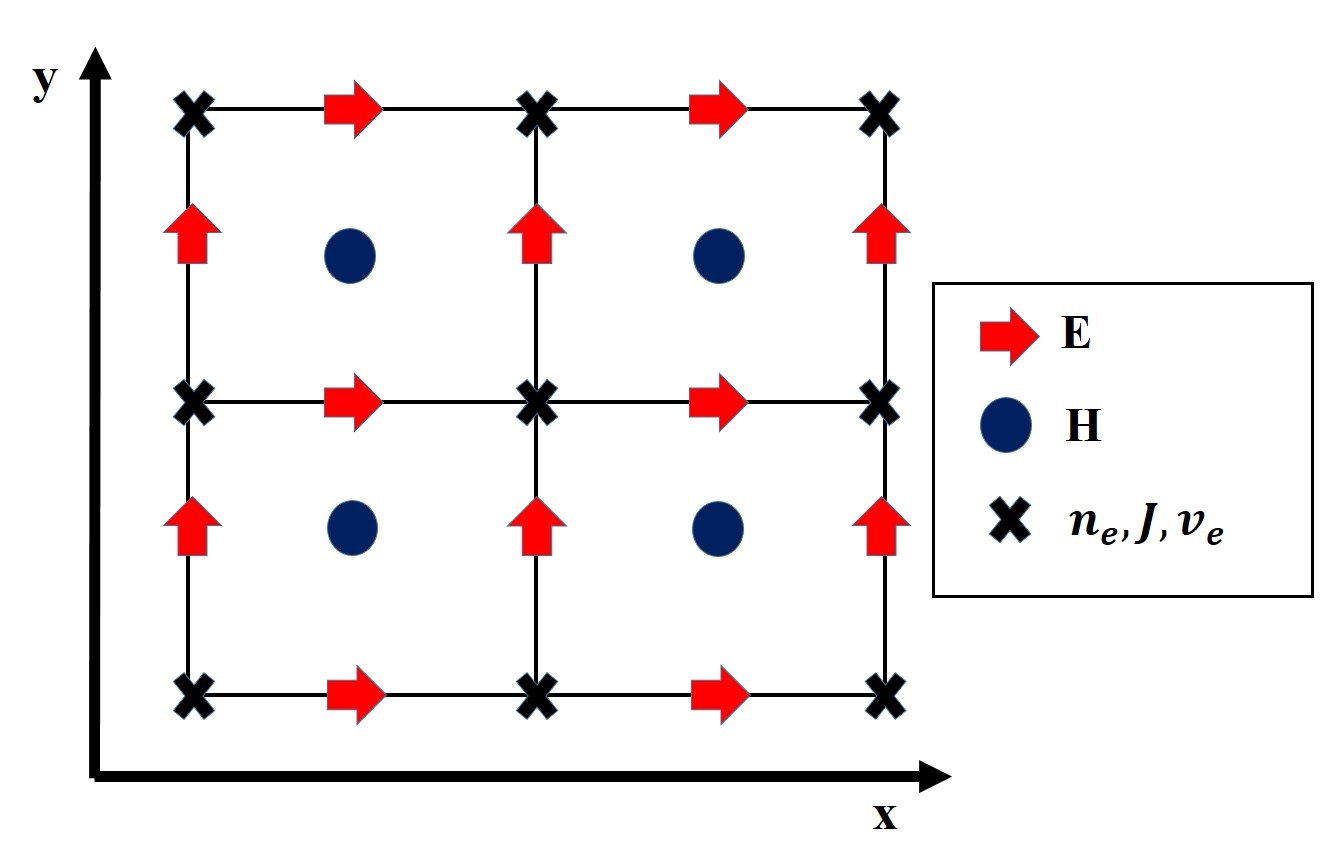}~
\caption{ 2D cartesian grid representation of FDTD (E and H-field) and plasma fluid model: plasma velocity ($v_{e}$) and current density ($J$) solver. The square $4 \times 4$ grid, each grid size ($\Delta$). The locations of E-field , H-field, plasma density ($n_{e}$), $J$ and $v_{e}$ are indicated. Proper spatial averaging required to evaluate the $n_{e}$, $J$ and $v_e$ at same locations of E-field.   %$\{(c_{kx},c_{ky})\in \mathbb{Q^{+}}\}$; length of the domain $L_{x}$ and $L_{y}$ are taken in terms of wavelength of the incident EM wave. The location $x_{0}$, $y_{0}$ is $0.5L_{x}$ and $0.5L_{y}$, respectively, and, $\{c_{kx},c_{ky}\}=\{1.5,1.5\}$, where $\lambda$ corresponds to freq = 1 GHz. The parameter space represents 1D plasma profile along the central x-axis ($x_{central}$) through the gaussian plasma (b) for different widths of gaussian function (S1 to S10) for a fixed peak plasma density,$n_{0}=10^{22}$ m$^{-3}$ (c) for a fixed width of gaussian functon having different peak plasma density. S1:$0.05\lambda$ (highest) to S10:$0.02\lambda$ (lowest). The den1S1: $n_{0}=10^{22}$ m$^{-3}$ to den6S1: $n_{0}=10^{21}$ m$^{-3}$
}\label{Fig:FDTDdicretization}
\end{figure}
%%%%%%%%%%%%%
For our 2-D simulation, discretized form of the system of Maxwell's equations , $E_{x}$, $E_{y}$, and $H_{z}$, as well as the charge particle velocity equation has been considered:

%\begin{subequations}\label{eq:5}
%\begin{gather}
%\frac{E_{x,i,j}^{n+1}-E_{x,i,j}^{n}}{\Delta t} 
%= \frac{1}{\epsilon_0}\lbrace\frac{H_{z,i,j}^{n-1/2}-H_{z,i,j-1}^{n-1/2}}{\Delta}\rbrace \tag{\theequation a}\\
%-\frac{1}{\epsilon_0}\lbrace\frac{J_{x,i,j}^{n+1}+J_{x,i,j}^{n}}{2}\rbrace \notag\\
%\frac{E_{y,i,j}^{n+1}-E_{y,i,j}^{n}}{\Delta t}
%= \frac{1}{\epsilon_0}\lbrace\frac{H_{z,i-1,j}^{n-1/2}-H_{z,i,j}^{n-1/2}}{\Delta}\rbrace \tag{\theequation b}\\
%- \frac{1}{\epsilon_0}\lbrace\frac{J_{y,i,j}^{n+1}+J_{y,i,j}^{n}}{2}\rbrace \notag\\
%\frac{H_{z,i,j}^{n+1/2}-H_{z,i,j}^{n-1/2}}{\Delta t} 
%= \frac{1}{\mu_0}\lbrace\frac{E_{x,i,j}^{n}-E_{x,i,j+1}^{n}}{\Delta}\tag{\theequation c}\\-\frac{E_{y,i+1,j}^{n}-E_{y,i,j}^{n}}{\Delta}\rbrace \notag\\
%\frac{v_{e_{x,i,j}}^{n+1}-v_{e_{x,i,j}}^{n}}{\Delta t} 
%= \frac{e}{m}\lbrace\frac{E_{{total}_{x,i,j}}^{n+1}+E_{{total}_{x,i,j}}^{n}}{\Delta}\rbrace \tag{\theequation d}\\
%-\nu_{m}\lbrace\frac{v_{e_{x,i,j}}^{n+1}+v_{e_{x,i,j}}^{n}}{2}\rbrace \notag\\
%\frac{v_{e_{y,i,j}}^{n+1}-v_{e_{y,i,j}}^{n}}{\Delta t} 
%= \frac{e}{m}\lbrace\frac{E_{{total}_{y,i,j}}^{n+1}+E_{{total}_{y,i,j}}^{n}}{\Delta}\rbrace \tag{\theequation e}\\
%-\nu_{m}\lbrace\frac{v_{e_{y,i,j}}^{n+1}+v_{e_{y,i,j}}^{n}}{2}\rbrace \notag
%\end{gather}
%\end{subequations}
\begin{subequations}\label{eq:5}
\begin{equation}
\begin{split}
   \frac{E_{x,i,j}^{n+1}-E_{x,i,j}^{n}}{\Delta t} &= 
   \frac{1}{\epsilon_0}\Biggl\{\Bigg(\frac{H_{z,i,j}^{n-1/2}-H_{z,i,j-1}^{n-1/2}}{\Delta_{y}}\Biggr) \\
   &\phantom{{}= \frac{1}{\epsilon_0}}- \Biggl(\frac{J_{x,i,j}^{n+1}+J_{x,i,j}^{n}}{2}\Biggr)\Biggr\}
 \end{split}
\end{equation}
\begin{equation}
\begin{split}
   \frac{E_{y,i,j}^{n+1}-E_{y,i,j}^{n}}{\Delta t} &= 
   \frac{1}{\epsilon_0}\Biggl\{\Bigg(\frac{H_{z,i-1,j}^{n-1/2}-H_{z,i,j}^{n-1/2}}{\Delta_{x}}\Biggr) \\
   &\phantom{{}= \frac{1}{\epsilon_0}}- \Biggl(\frac{J_{y,i,j}^{n+1}+J_{y,i,j}^{n}}{2}\Biggr)\Biggr\}
\end{split}
\end{equation}
\begin{equation}
\begin{split}
   \frac{H_{z,i,j}^{n+1/2}-H_{z,i,j}^{n-1/2}}{\Delta t}  &= 
   \frac{1}{\mu_0}\Biggl\{\Bigg(\frac{E_{x,i,j}^{n}-E_{x,i,j+1}^{n}}{\Delta_{y}}\Biggr) \\
   &\phantom{{}= \frac{1}{\mu_0}}- \Biggl(\frac{E_{y,i+1,j}^{n}-E_{y,i,j}^{n}}{\Delta_{x}}\Biggr)\Biggr\}
 \end{split}
\end{equation}
\begin{equation}
\begin{split}
   \frac{v_{e_{x,i,j}}^{n+1}-v_{e_{x,i,j}}^{n}}{\Delta t}  &= 
   \Biggl\{\frac{e}{m_{e}}\Bigg(\frac{E_{{total}_{x,i,j}}^{n+1}+E_{{total}_{x,i,j}}^{n}}{2}\Biggr) \\
   &\phantom{{}= }- \nu_{m}\Biggl(\frac{v_{e_{x,i,j}}^{n+1}+v_{e_{x,i,j}}^{n}}{2}\Biggr)\Biggr\}
 \end{split}
\end{equation}
\begin{equation}
\begin{split}
   \frac{v_{e_{y,i,j}}^{n+1}-v_{e_{y,i,j}}^{n}}{\Delta t}  &= 
   \Biggl\{\frac{e}{m_{e}}\Bigg(\frac{E_{{total}_{y,i,j}}^{n+1}+E_{{total}_{y,i,j}}^{n}}{2}\Biggr) \\
   &\phantom{{}= }- \nu_{m}\Biggl(\frac{v_{e_{y,i,j}}^{n+1}+v_{e_{y,i,j}}^{n}}{2}\Biggr)\Biggr\}
 \end{split}
\end{equation}
\end{subequations}

%\begin{flalign}
%\frac{E_{x,i,j}^{n+1}-E_{x,i,j}^{n}}{\Delta t}
%= \frac{1}{\epsilon_0}\left({\frac{H_{z,i,j}^{n-1/2}-H_{z,i,j-1}^{n-1/2}}{\Delta}}\right)\nonumber\\ -\frac{1}{\epsilon_0}\left({\frac{J_{x,i,j}^{n+1}+J_{x,i,j}^{n}}{2}}\right) 
%\tag{6a}
%\end{flalign}
The grid size ($\Delta_{x}=\Delta_{y}=\Delta$) is decided based on the  minimum of EM wave frequency (wavelength) and gradient of plasma density to be resolved \cite{BHASKAR2007}. The time step for each iteration ($\Delta t$), satisfies CFL criteria for a stable FDTD solution. The root mean square (RMS) of the E-field obtained from microwave-plasma scattering have been used for the data generation purpose. The $E_{rms}$, is the time-averaged E-field over one EM wave period , $E_{rms}=\left(\frac{1}{N}\sum_{i=1}^{N}{ E_{total}^{2}}\right)^{1/2}$, where $i$ is the number of iteration going upto $N$, corresponds to $1$ EM wave period and $E_{total}$, total E-field, $E_{total}= E_{scattered}+E_{incident}$ \cite{Kunz-1993}.
%\textcolor{red}{mention here how ERMS is calculated - based on which image is generated}The total electric (E) field, $E_{total}=E_{incident}+E_{scattered}$. 
%The plasma current density ($J$), $\left(J=-e\:n_e\: v_e\right)$ in (A m$^{-2}$) . 
%The sharp variation at the plasma-air interface results in sharp gradients in the E-field at the boundary that requires to be properly resolved using discrete space and time steps. 
%For better accuracy in the evaluated Fields, a fully uniform fine meshing has been considered in the plane of the domain consisting of the E-field parallel to the domain. The chosen grid size is 512 cells in a given wavelength of the microwave E-field \cite{BCHAUDHURY2018} referred here as $N_{\lambda}$. 
%The stringent requirements results in trade-off between accuracy and the total simulation time. To avoid such challenges an  optimal mesh refinement algorithm based on sub-gridding technique has been implemented~\cite{Pratik2020} to obtain fast and near accurate results. Two grids, coarse and fine grids with a refinement factor of 2, have been used in the computational domain where the EM-field and plasma density values are updated.
The solution to the above discretized coupled microwave plasma fluid model, represented by Eq. \ref{eq:5} (a-e) can be alternatively solved using the deep learning based technique which will be discussed in the next section.
\section{Proposed Deep Learning Architecture} \label{sec3}

\begin{figure}[!htbp]
\centering
\includegraphics[width = 0.95\columnwidth]{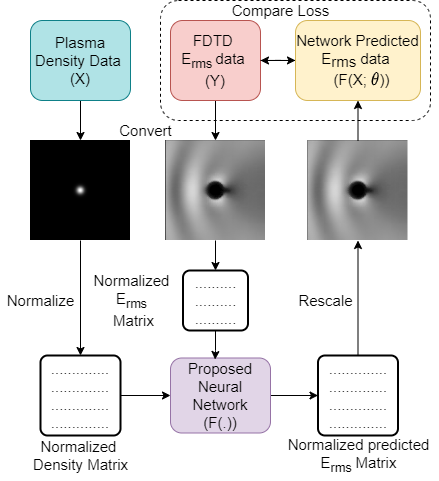}~
\caption{Flowchart for training the proposed architecture}\label{Fig:pipeline}
\end{figure}
%%%%%%%%%%%%%%%%%%%%
%%%%% Figure4 %%%%%%%%%%%%
\begin{figure*}[ht]
\centering
\includegraphics[width = \textwidth]{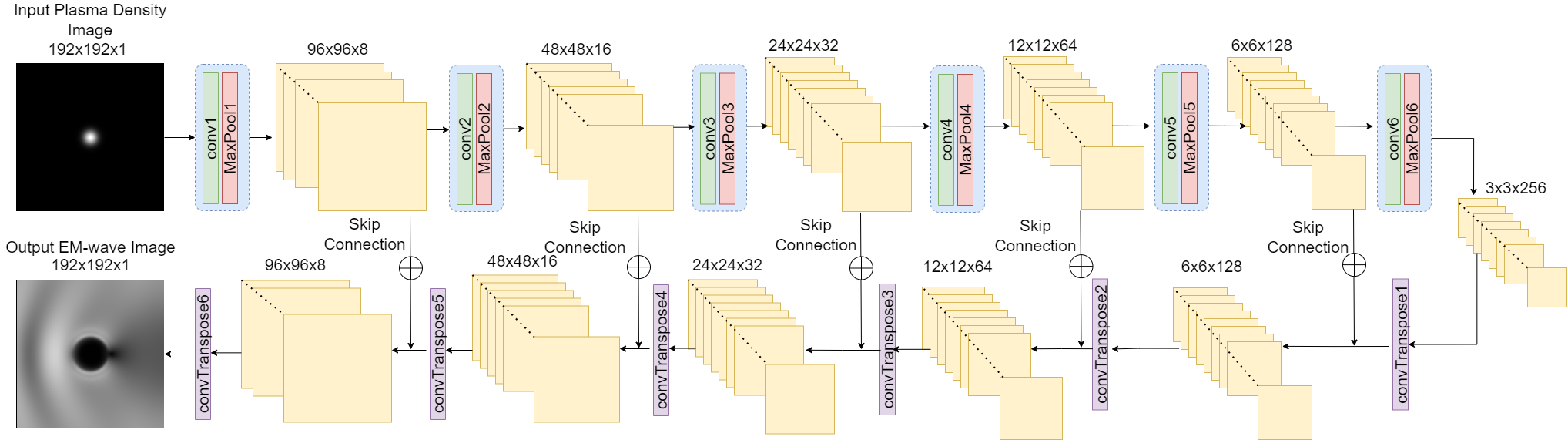}~
\caption{The proposed architecture having encoder, decoder and skip connections for EM-wave scattered by plasma density prediction }\label{Fig:DL_net}
\end{figure*}
\par
%%%%%%%%%%%%%%%%%%
In this section, a deep learning based architecture is proposed for solving the microwave-plasma scattering problem given the plasma density and an EM wave of fixed frequency. A general flowchart followed for carrying out the experiments in the next section is shown in Fig. \ref{Fig:pipeline} and is explained as follows: Deep learning models are trained on image datasets, therefore the data obtained from the FDTD solver needs to be converted into images. To generate the data and to train the deep learning model, the plasma density and $E_{rms}$ data from the 2D FDTD based simulations is normalized between 0-1 using the maximum value of the plasma density and $E_{rms}$ obtained from the complete dataset respectively. Both of these maximum values for plasma density (in $m^{-3}$) and $E_{rms}$ (in $V/m$) are saved and used for scaling up the normalized output which is generated by the trained neural network. The normalized dataset are scaled in the range (0-255) and gray-scale images are generated for training the proposed deep learning network.
 
% The training data is separated from the test data, which is generated from the images by converting them into numpy arrays. These are again normalized (0-1) before training on the deep learning model. 
\par
The generated pair of plasma density ($X$) and $E_{rms}$ ($Y$) images are then used to train the proposed deep learning model. The model is then evaluated on the testing data. Plasma density image $X$ is given as input to the trained network which outputs the predicted $E_{rms}$ image (dentoed by $F(X;\theta)$, where $F$ represents the deep learning model and $\theta$ is the trained model weight matrix. The predicted $E_{rms}$ image is converted to the physical $E_{rms}$ values (in V/m) by scaling the normalized output by the global maximum of the dataset as discussed earlier. The predicted $E_{rms}$ values from the deep learning model are then compared with the actual $E_{rms}$ values from the 2D FDTD based computational solver.
\par
The proposed architecture is a CNN-based UNet\cite{b2ORonneberger2015} where the input to the network is the single-channeled, gray-scaled, normalized image of plasma density $X$ and the output to the network is the corresponding single-channeled, gray-scaled, normalized image of the $E_{rms}$ data.  The architecture of the network is shown in Fig. \ref{Fig:DL_net}. It can be seen that the grayscale plasma density image is given as an input to the network.
The model consists of an encoder and decoder like structures. The encoder consists of series of convolutional and max pooling layers which learns the features from the image and reduces the dimensions in each layer. It helps the network to learn  training weights and the reduction in image dimension decreases the complexity of the model. There are six encoder units each having a convolutional layer with $n$ filters where $n$ is twice the number of filters than the previous unit having $3 \times 3$ kernel size.  The output of each layer is followed by the ReLU activation function. 
\par
Correspondingly, there are six decoder units with each unit having the transposed convolution operation layer with kernel size of $2 \times 2$ followed by the ReLU activation function. The decoder layer will upsample the features to construct the output image of the network. The input to the decoder layer is connected directly to the output of the encoder. Each layer of the decoder is connected to the corresponding encoder unit's output using a skip connection as shown in the Fig. \ref{Fig:DL_net}. The skip connections are implemented by concatenating the output of one layer to the other layer to which it is connected. The output of the final decoder unit is the predicted $E_{rms}$ image from the proposed architecture.

\section{Experimental Work}

%%%% Figure6 %%%%%%%%%%
\begin{figure}[h]
\centering
\includegraphics[width=1.0\textwidth, height=0.74\textwidth]{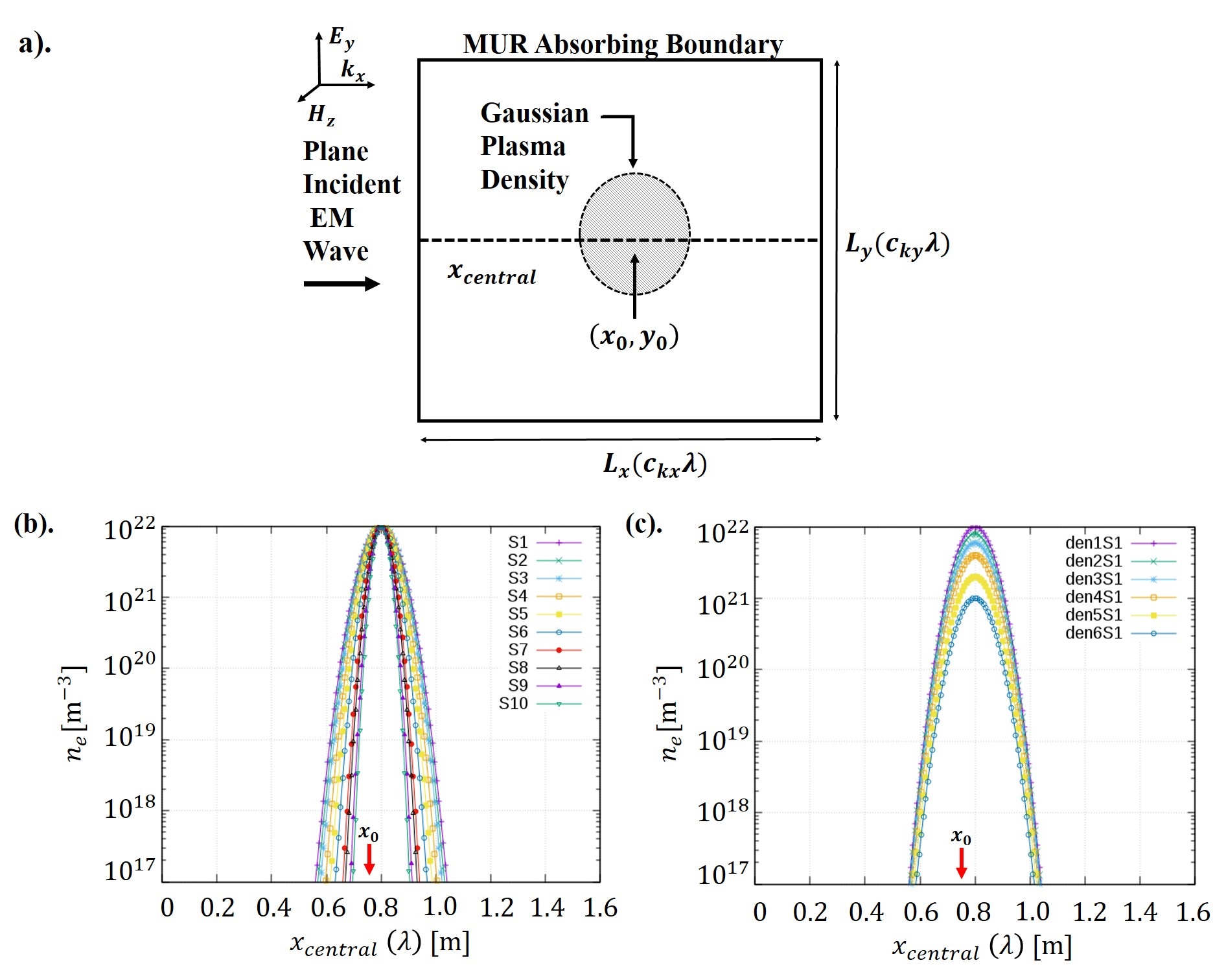}~
\caption{(a). The schematic representation of square computational domain, $\{(c_{kx},c_{ky})\in \mathbb{Q^{+}}\}$; length of the domain $L_{x}$ and $L_{y}$ are taken in terms of wavelength of the incident EM wave. The location $x_{0}$, $y_{0}$ is $0.5L_{x}$ and $0.5L_{y}$, respectively, and, $\{c_{kx},c_{ky}\}=\{1.5,1.5\}$, where $\lambda$ corresponds to freq = 1 GHz. The parameter space can be varied by changing two parameters of the 2D Gaussian profile - width and peak density. (b) plasma profile along the central x-axis ($x_{central}$) for different widths of Gaussian (S1:$0.05\lambda$ (highest) to S10:$0.02\lambda$ (lowest)) for a fixed peak plasma density,$n_{0}=10^{22}$ m$^{-3}$ (c) different peak plasma densities for a fixed width of Gaussian. The den1S1: $n_{0}=10^{22}$ m$^{-3}$ to den6S1: $n_{0}=10^{21}$ m$^{-3}$}\label{Fig:gaussianwidth}
\end{figure}
%%%%%%%%%%%%%%%%%%%%%%
\subsection{Dataset generation}
As discussed in Sec. \ref{sec3} that deep learning architectures require training data. Therefore, in this section a discussion about how the dataset is generated is carried out. Fig. \ref{Fig:gaussianwidth}(a) provides a phenomenological picture of the 2D problem we are trying to simulate. Let us consider a linearly polarized plane EM wave propagating in air plasma in the X direction. The simulation plane XY contains electric field E and the wave
vector k parallel to the X direction. The magnetic
field H is in the YZ plane perpendicular to the X direction.
This is equivalent to a Y-polarized, X-directed wave.
A 2D gaussian plasma density has been considered given by, $n_{e}(x,y)=n_{0}exp(-(\{x-x_{0}\}^{2}/\sigma_{x}^2+\{y-y_{0}\}^{2}/\sigma_{y}^2))$, where $x_{0}$ and $y_{0}$ are the center of the plasma peak density ($n_0$) and the spread of the plasma is controlled by the plasma width $\sigma_{x}$ and $\sigma_{y}$, here $\sigma_{x}=\sigma_{y}=\text{S}$. Different Gaussian plasma profile can be defined by tuning the two important parameters, the width of the Gaussian, $\text{S}$, and the peak plasma density, $n_{0}$. Some representative plasma profiles for generating the dataset (pair of plasma density and $E_{rms}$ field) to study the EM-plasma interaction are shown in Figs. \ref{Fig:gaussianwidth}(b)-(c).
%%%%%%%%%%%%%%%%%%%%%%%%%%%%%%%%%%%%%%%
% [width = 0.7\textwidth,height=0.33\textwidth]
\begin{figure*}[]
\centering
\includegraphics[width = 0.84\textwidth,height=0.4\textwidth]{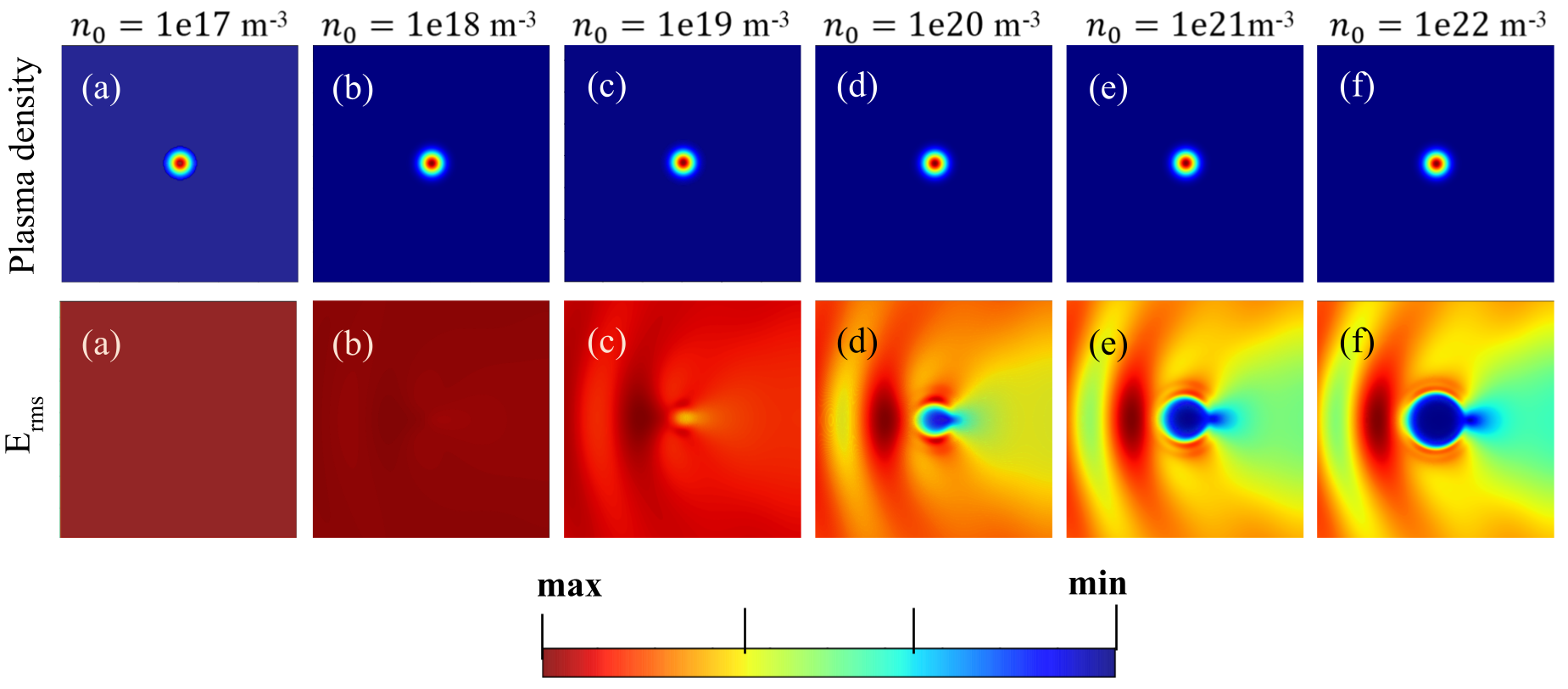}~
\caption{(a-f) Generated dataset of Plasma Density and corresponding scattered $E_{rms}$ for varying peak density. The colorbar represents the maxima and minima corresponds to both plasma density and the $E_{rms}$. The maxima for plasma density is indicated by $n_{0}$ and minima is 0. For $E_{rms}$, maxima are 7.07, 7.15, 7.74, 9.37, 10.36 and 10.47 V/m respectively and minima is 0. The skin depth of microwave into plasma profile reduces as $n_{0}$ increases indicated by visibility of exact scatterer dimension (2D Gaussian profile) from $E_{rms}$ plot (a) to (f).}\label{Fig:Dataset_ex}
\end{figure*}
%%%%%%%%%%%%%%%%%%%%%%%%%%%%%%%%%%%%%%%%%%%%%%%%5
The dataset to train the network is prepared by keeping the frequency of the incident wave fixed and the shape of the 2D gaussian plasma profile is varied. The size of the computational domain is 1.5$\lambda$ x 1.5$\lambda$ as per the setup shown in Fig. \ref{Fig:gaussianwidth}. For each instance of the profile, the data file for the plasma density and the corresponding 2D scattered EM wave data ($E_{rms}$) is generated via our in-house developed FDTD computational solver \cite{Pratik2020}. The generated data is in the form of a 2D grid and is visualized in the Fig. \ref{Fig:Dataset_ex} for varying peak density. It can be observed that the EM wave transmits through in the case of lower peak densities while it gets reflected in the case of higher peak densities.

% \subsection{Parameter Selection}
% Mihir all the parameter selection criteria has to be added here

\subsection{Training Details}
\begin{figure}[!htbp]
\centering
\includegraphics[width = 0.45\textwidth]{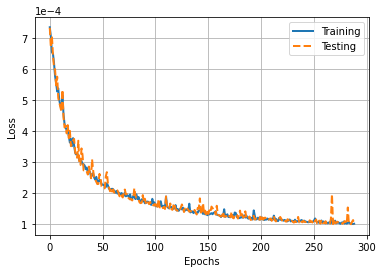}~
\caption{Mean squared error loss in training the model} \label{Fig:loss}
\end{figure}

The proposed deep learning network is trained on the pair of the generated gray-scale images of the dataset. The plasma density image $X$ is given as input to the network, and the network learns its parameters by minimizing the loss between the actual $E_{rms}$ image denoted as $Y$ and the output of the network which is the predicted $E_{rms}$ image denoted as $F(X;\theta)$. The loss function for training the architecture is given as follows:
\begin{equation}
    L(\theta) = \frac{1}{M} \mathlarger{\sum}_{i=1}^M \lVert F(X;\theta) - Y \rVert_{2}^{2} + \lambda \mathlarger{\sum}_{j=1}^l \lVert W_j  \rVert_{1}
\end{equation}
where $M$ is the total number of training images, $\theta$ is the network weight parameter matrix, $l$ is the total number of kernels used and $W_j$ is the weight of the $j^{th}$ kernel. The loss is minimized using the Adam optimizer\cite{Kingma} with learning rate $\eta = 1e-3$, $\beta_1 = 0.9$ (the exponential decay rate for first-order moment estimates), $\beta_1 = 0.999$ (the exponential decay rate for second-order moment estimates) and $\epsilon = 1e-7$. The kernel weights matrix for the convolution and transposed convolution layers are initialized with Glorot-uniform which draws samples from a uniform distribution. Here, L1 regularization is used to overcome the problem of overfitting with $\lambda = 1e-7$. There are 611,833 trainable parameters in the proposed deep learning model with 6 convolutional encoders, 6 convolutional transposed decoders and 5 skip connections in between. The network is trained on NVIDIA Tesla K40c GPU using Keras API with Tensorflow running in backend. The loss while training the model for 300 epochs is shown in Fig. \ref{Fig:loss}. It can be observed that training and testing losses decreases with each epoch indicating that our network is learning.
%%%%%% Figure 5 %%%%
%%%%%%%%%%%%%%%%%%%%

%%%%%%%%%%%%%%%%%%%%%
\subsection{Performance Comparison Metrics}
To evaluate the performance of the proposed architecture in terms of predicting the image as well as the actual physical values of the $E_{rms}$ data (in V/m), we have used existing metrics that are discussed in this section. In order to compare the quality of image reconstruction by the deep learning model with respect to the actual $E_{rms}$ image, we use Structural Similarity Index Metric (SSIM)\cite{SSIM}, where the image degradation is perceived change in structural information. For two images $x$ and $y$ of size $N \times N$, the similarity measure is given by:
\begin{equation}
    SSIM = \frac{(2 \mu_x \mu_y + (k_1L)^2)(2\sigma_{xy} + (k_2L)^2)}{(\mu_x^2 + \mu_y^2 + (k_1L)^2)(\sigma_x^2 + \sigma_y^2 + (k_2L)^2)}
\end{equation}
where $\mu$ denotes the mean, $\sigma^2$ denotes the variance, $L$ is the dynamic range of pixel values and $k_1 = 0.01$ and $k_2 = 0.03$. Values of SSIM index are from 0 to 1. Closer the value of the SSIM metric to 1, better the quality of image reconstruction.
\par

To evaluate the performance of the model in predicting the $E_{rms}$ values (in $V/m$) in comparison the values obtained from the 2D FDTD based solver, we use two metrics. The first metric is the average of the percentage error over all the $E_{rms}$ values on the 2D grid. Let $A_{ij}$ and $B_{ij}$ denote the $E_{rms}$ values obtained from the 2D FDTD solver and DL based approach at $(i,j)^{th}$ point on a $N \times N$ 2D grid respectively. The average percentage error is given by
\begin{equation}
    \mbox{Avg. percent error} = \mathlarger{\sum}_{i=1}^N \mathlarger{\sum}_{j=1}^N \frac{|B_{ij} - A{ij}|}{A_{ij}}
\end{equation}
\par
The second metric is the mean squared error (MSE) which is defined as
\begin{equation}
    MSE = \frac{1}{N^2} \mathlarger{\sum}_{i=1}^N \mathlarger{\sum}_{j=1}^N (B_{ij} - A{ij})^2
\end{equation}

\subsection{Experimental Results}
In the microwave-plasma interaction study, the 2D FDTD based method discretizes the 2D computational domain of size $1.5\lambda\times1.5\lambda$ using Yee approximation \cite{Yee-1966}. The data provided to the proposed DL based model is such that number of grid points per wavelength ($\lambda$) of the EM wave is $128$ resulting into $192\times192$ grid points in XY plane to accurately resolve the gradients in the E-field and the plasma density. 
%the computational domain is a square with dimensions in terms of the wavelength ($\lambda$) of the incident plane EM wave is $1.5\lambda$ by $1.5\lambda$. The square is uniformly discretized into fine mesh throughout the domain to capture the EM-plasma transmission, scattering and absorption both inside and at the plasma-free space boundaries. The accuracy of the simulated results is highest when total cells per wavelength ($\lambda$) is 512 \cite{BCHAUDHURY2018} for E-fields in the plane of the square. increases without increasing amount of computation that would have resulted from a uniform fine mesh in entire square. The grid size ($\Delta_{x}=\Delta_{y}=\Delta$) is $0.5$ mm similarly, based on the CFL criteria, the time step ($\Delta t$) is $0.95$ ps.
In the simulation, the plane EM wave having the amplitude of $10$ V/m is incident from the left hand side of the domain as shown in Fig. \ref{Fig:gaussianwidth}. The frequency of the EM wave is $1$ GHz. %The details of the hardware architecture for the computer on which the FDTD based EM-plasma interaction data generated is as follows, Intel Xeon CPU E5-2640 V3 @ 2.60 GHz on x86\_64 architecture. 
\par

% The model is trained on the dataset with varying the range of the densities from $n_e = 1e18 m^{-3}$ to $n_e = 1e22 m^{-3}$. While compiling the model, adam optimizer \cite{Adam} is used. The loss function used here is mean squared error (MSE) with L1 regularization which is given by:
% \begin{equation}\label{eq6}
% \centering
%     MSE = \frac{1}{n} \sum_{i=1}^n (y_i - \hat{y_i})^2 + \lambda \sum_{i=1}^n (y_i - \hat{y_i})
% \end{equation}
% where $y_i$ is the actual value and $\hat{y_i}$ is the predicted value. The model is left to be trained for 200 epochs. The loss plot for the first 30 epochs is shown in \ref{Fig:loss}.

% \begin{figure}[htbp]
% \centering

% \includegraphics[width=0.5\textwidth, height=0.2\textwidth]{Figure_1.png}~
% \caption{Output image of network compared with actual image}\label{Fig:diffim}
% \end{figure}
% \begin{figure*}[]
% \centering
% \includegraphics[width = \columnwidth]{results_plot_h.png}~
% \caption{Proposed deep learning architecture for EM-Wave scattering prediction with 6 encoder, 6 decoder units with 5 skip connections}\label{Fig:DL_net}
% \end{figure*}

%%Figure 7%%%
\begin{figure*}[!htbp]
\centering
\includegraphics[width =\textwidth]{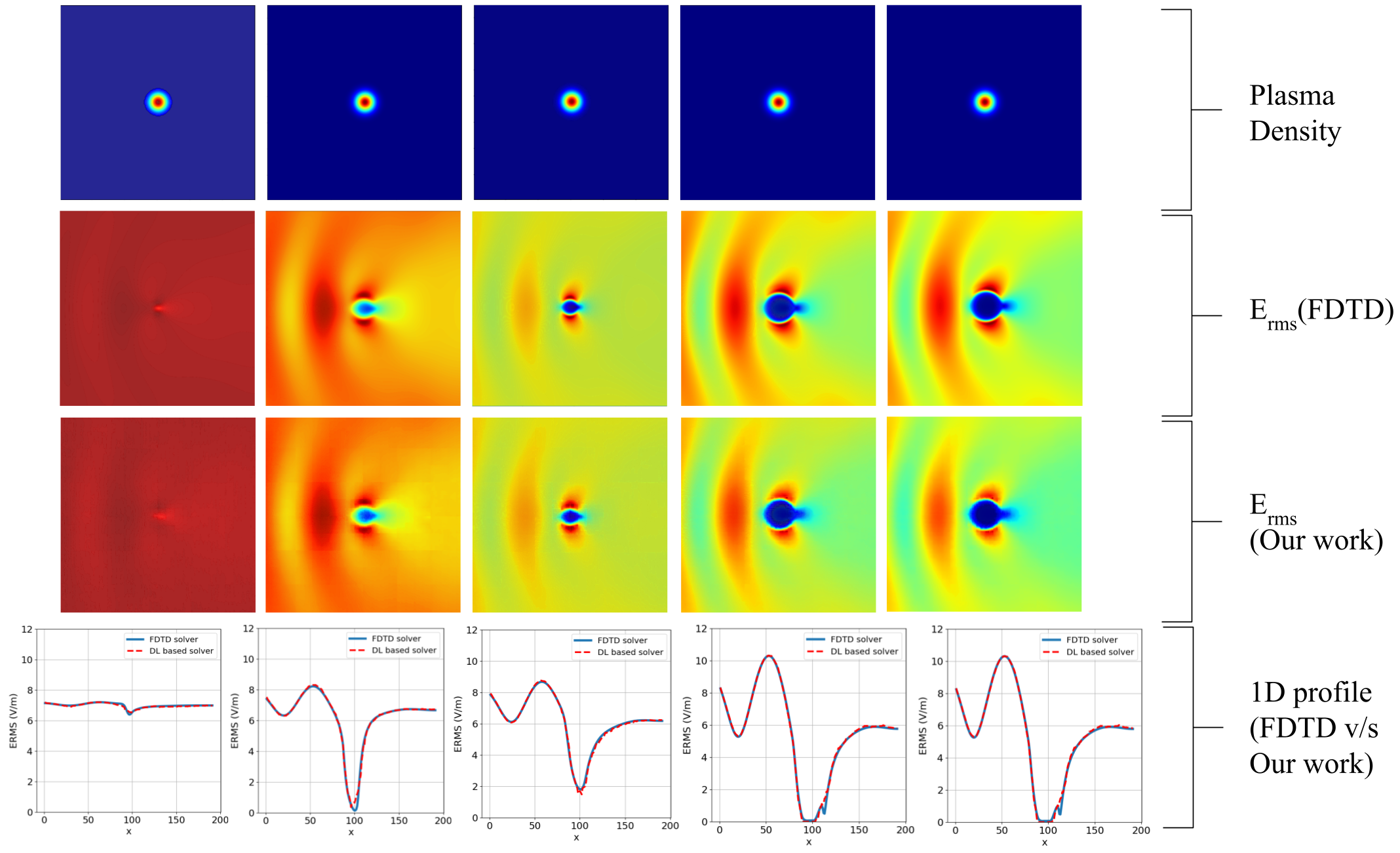}~
\caption{Comparative study and results of example cases; Row 1: Input plasma density (increaing order from left to right) image; Row 2: Scattering ($E_{rms}$) pattern obtained from FDTD solver; Row 3: Output $E_{rms}$ image from the proposed deep learning based architecture; Row 4: Comparison of 1D $E_{rms}$ across the central x-axis of the computational domain predicted from our work and FDTD based solver.}\label{Fig:res_app}
\end{figure*}
% \begin{figure}[htbp]
% \centering
% \includegraphics[width = \columnwidth]{results_plot_a.png}~
% \caption{Actual and predicted images of the EM-Wave scattering and comparison of prediction v/s traditional solver in 1D cross section across central axis for a high density profile}\label{Fig:res_a}
% \end{figure}
% %%Figure 8%%%
% \begin{figure}[htbp]
% \centering
% \includegraphics[width = \columnwidth]{results_plot_b.png}~
% \caption{Actual and predicted images of the EM-Wave scattering and comparison of prediction v/s traditional solver in 1D cross section across central axis for a medium density profile}\label{Fig:res_b}
% \end{figure}
% %%Figure 9%%%
% \begin{figure}[htbp]
% \centering
% \includegraphics[width = \columnwidth]{results_plot_c.png}~
% \caption{Actual and predicted images of the EM-Wave scattering and comparison of prediction v/s traditional solver in 1D cross section across central axis for a low density profile}\label{Fig:res_c}
% \end{figure}
%%%%%%%%%%%%%
% \begin{figure}[htbp]
% \centering
% \includegraphics[width = \columnwidth]{1D_pred.png}~
% \caption{Output image of network compared with actual image}\label{Fig:1d}
% \end{figure}
The data used in training the network is varied by changing the peak plasma density from $n_0 = 1e21 m^{-3} \rightarrow 1e22 m^{-3}$ to $n_0 = 1e17 m^{-3} \rightarrow 1e22 m^{-3}$ gradually. The results on the test cases with varying dataset size in training are shown in Table ~\ref{tab1}. It can be observed that the average SSIM index increases with the increase of the dataset range. Average percent error is observed to be less than 2\%. The results obtained from the FDTD based solver and DL based approach for different plasma densities have been compared in Fig. \ref{Fig:res_app}. The first row in Fig. \ref{Fig:res_app} represents density profiles changing from underdense (leftmost) plasma to overdense plasma (rightmost). The second row represents the scattered EM wave pattern obtained using the conventional FDTD based solution. Third row shows the generated images from our proposed architecture and it can be observed that the scattering patterns closely matches with that of the FDTD based solver data shown in second row. Fourth row shows the quantitative comparison of the scattered EM wave values shown in second and third rows. It can be observed that the intensity values obtained by taking 1D profile across the central x-axis closely matches with each other (FDTD vs our proposed approach).
%%%% Table 1%%%%%%%%%%
\begin{table}[htbp]
\centering
\caption{\label{tab1} EM-Wave scattering predicted data comparison with the actual data}
\begin{tabular}{|l|l|l|l|}
\hline
\begin{tabular}[c]{@{}l@{}}Dataset\\ Range(in density $m^{-3}$)\end{tabular} & \begin{tabular}[c]{@{}l@{}}Avg. SSIM\\ (image)\end{tabular} & \begin{tabular}[c]{@{}l@{}}Avg. percent error\\ (physcial)\end{tabular} & \begin{tabular}[c]{@{}l@{}}Avg. MSE\\ (physical)\end{tabular} \\ \hline
1e21 - 1e22                                                         & 0.9894                                                      & 1.8751\%                                                                & 0.00857                                                       \\ \hline
1e20 - 1e22                                                         & 0..9932                                                     & 1.1017\%                                                                & 0.00467                                                       \\ \hline
1e19 - 1e22                                                         & 0.9935                                                      & 1.3389\%                                                                & 0.00722                                                       \\ \hline
1e18 - 1e22                                                         & 0.9946                                                      & 1.0172\%                                                                & 0.00613                                                       \\ \hline
1e17 - 1e22         &  0.9955     &  0.9036\%     & 0.00564 \\ \hline
\end{tabular}
\medskip
\end{table}
\par
The aforementioned microwave-plasma interaction study in a $1.5\lambda\times1.5\lambda$ computational domain using the FDTD based technique with 128 cells per $\lambda$ takes approximately 18 seconds on Intel Xeon CPU E5-2640 V3 @ 2.60 GHz with x86\_64 architecture for a physical time duration of 15 wave periods. The stable scattering pattern is obtained after the EM wave has attained steady state condition. 
%The wall-time taken by the trained model to output the image and rescale it to final $E_{rms}$ values and write the final output in individual .npy files for every test case using $process\_time()$ function.
%It took 8.998 seconds to run the 156 test cases. 
Whereas, it takes $0.0576$ seconds on average per test case for the DL based approach executed on the Intel(R) Xeon(R) CPU @ 2.20GHz with x86\_64 architecture. \\
\indent
For the $1.5\lambda\times1.5\lambda$ problem size, we observe a speedup of around 350 times by using the DL based approach compared to the FDTD based technique. However, the computational time complexity of 2D FDTD based solver is $O(n^3)$, and if the problem size changes from $1.5\lambda\times1.5\lambda$ to $3\lambda\times3\lambda$ and finally to $6\lambda\times6\lambda$, the execution time is around 93 and 744 seconds respectively. However, in the case of DL based approach, it will be much smaller.
\subsection{Ablation Study} 
In order to explore the effect of changing the model parameters on the predicted $E_{rms}$ values given the plasma density, the ablation studies on the effects of skip connections, number of encoder-decoder units and method of upsampling are discussed below. In these experiments, all the models are trained (on the dataset having range $1e17 m^{-3} - 1e22 m^{-3}$) on the same training and testing images.

\subsubsection{Effect of encoder-decoder units} 
In the proposed architecture, there are six encoder and decoder units. In this study, we perform additional experiments with four and five encoder-decoder units in the proposed architecture. The results are shown in Table \ref{Tab1} where it can be observed that adding more units will improve the results. For six unit pairs, we obtain an average SSIM value of more than $0.99$ which is a good indicator for image reconstruction by the network. Similar observation is seen for average percentage and mean square error (MSE) metrics. Further, adding more encoder-decoder units will just increase the computational cost.
\begin{table}[!htbp]
\centering
\caption{Impact of varying the number of encoder-decoder units}
\begin{tabular}{|l|l|l|l|}
\hline
\begin{tabular}{@{}l@{}}No. of encoder- \\decoder units\end{tabular}                                      & Avg. SSIM & Avg. Percent error & Avg. MSE \\
\hline
4 units          & 0.95881   & 4.919\%            & 0.29512  \\
\hline
5 units           & 0.97629   & 3.164\%            & 0.09468  \\
\hline
6 units           & \textbf{0.99574}   & \textbf{0.897\%}            & \textbf{0.00628}\\   
\hline
\end{tabular}
\medskip
\label{Tab1}
\end{table}

\subsubsection{Effect of skip connections} 
 In the proposed model, there are five skip connections which connects the output of first five encoder and decoder units. For a second case, all the skip connections are removed. In the third case, there are two skip connections between the output of second and fourth encoder and decoder unit. We found the best results in the third case when there are two alternate skip connections as shown in the Table \ref{Tab2} although the difference is much less. 
 
\begin{table}[!htbp]
\centering
\caption{Impact of varying the skip connections}
\begin{tabular}{|l|l|l|l|}
\hline
\begin{tabular}{@{}l@{}}No. of \\skip connections(SC)\end{tabular}                                      & Avg. SSIM & Avg. Percent error & Avg. MSE \\
\hline
All five SC           & 0.99574   & 0.897\%            & 0.00628  \\
\hline
Without SC  & 0.99533   & 0.955\%            & 0.00805  \\
\hline
With two alternate SC         & \textbf{0.99601}   & \textbf{0.857\%}            & \textbf{0.00600}\\   
\hline
\end{tabular}
\medskip
\label{Tab2}
\end{table}

\begin{figure*}[!htbp]
\centering
\includegraphics[width=\textwidth]{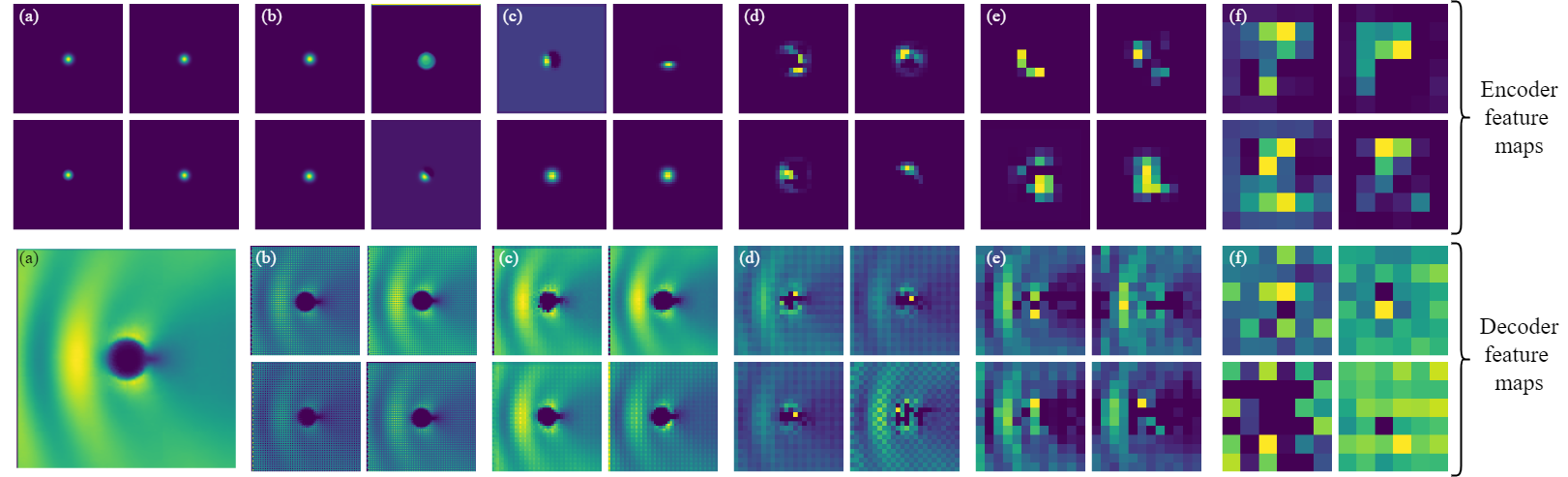}~
\caption{Row 1: Feature maps of the output of the six encoder units; Row 2: Feature maps of the output of the six decoder units. (a) is the final output $E_{rms}$ image of the network; The corresponding (a)-(f) pairs of the encoder-decoder pairs give the feature output in same spatial dimensions. (For representational purpose only four feature maps for a convolution unit are shown. Total number of feature maps for a unit is equal to the number of filters used in the convolution or transposed convolution layer.)}\label{Fig:fmap}
\end{figure*}

\subsubsection{Effect of up-sampling method} 
The proposed architecture uses the transposed convolution operation in the decoder units. The transposed convolution operation can be replaced by a structure consisting of an up-sampling unit followed by a convolution unit. Unlike the transpose convolution which is trainable, the up-sampling layers follow a interpolation scheme which increases the dimension of the input. The combination of up-sampling with convolution is considered equivalent to the transposed convolution \cite{Dumoulin}. We use the bilinear and nearest neighbour in this study and observe that using the up-sampling with nearest neighbourhood interpolation followed by convolution layers give the best results as shown in the Table \ref{Tab3}.
\begin{table}[!htbp]
\centering
\caption{Impact of using various up-sampling methods}
\begin{tabular}{|l|l|l|l|}
\hline
\begin{tabular}{@{}l@{}}Up-sampling \\Methods\end{tabular}                                    & Avg. SSIM & Avg. Percent error & Avg. MSE \\
\hline
\begin{tabular}{@{}l@{}}Transposed \\Convolution\end{tabular}             & 0.99574   & 0.897\%            & 0.00628  \\
\hline
\begin{tabular}{@{}l@{}}Bilinear \\ interpolation\end{tabular}             & 0.99568   & 0.960\%            & 0.00749  \\
\hline
\begin{tabular}{@{}l@{}}Nearest neighbour \\ interpolation\end{tabular}             & \textbf{0.99637}   & \textbf{0.857\%}            & \textbf{0.00579}\\   
\hline
\end{tabular}
\medskip
\label{Tab3}
\end{table}
\par

\subsection{Network visualization using feature maps}
Fig. \ref{Fig:fmap} shows the feature maps (outputs of the convolution units) for an example case which helps us visualize the way the network actually learns the features. It can be observed that the encoder section of the network (shown in row 1 in Fig. \ref{Fig:fmap}) learns the high level features initially followed by learning low level features (edges in the image) later. While the decoder of the network (shown in row 2 in Fig. \ref{Fig:fmap}) learns the low level features first followed by  learning the high level features. The encoder down-samples the image from (a) to (f), while the decoder up-samples the image from (f) to (a).  It can be observed that the decoder learns the scattering pattern unlike the encoder which just learns the structural features and down-samples the image resolution to give the structural information of the plasma density profile.

\section{Conclusion}
This work presents a CNN-based deep learning model, inspired from UNet with series of encoder and decoder units with skip connections, for the simulation of microwave plasma interaction. The scattering of a plane EM wave, with fixed frequency and amplitude, incident on a plasma medium with different Gaussian density profiles have been considered. The training data associated with microwave-plasma interaction have been generated using 2D-FDTD based simulations. The trained deep learning model is then used to reproduce the scattered EM wave from the plasma with average percent error margin of less than 2\%. The results obtained from the network have been evaluated using various metrics such as SSIM and MSE. Ablation studies along with network visualization using feature maps has also been discussed.
This work can be further expanded by training the network on various shapes of plasma profile. The deep learning technique proposed in this work is significantly fast compared to the existing FDTD based computational techniques.

\section*{Acknowledgment}
P. Ghosh would like to thank the DST, Gov. of India, for research fellowship received under DST-SERB project (Project No. - CRG/2018/003511).

\bibliographystyle{abbrv}
\bibliography{references}

\end{document}